\documentclass[a4paper,twoside]{article}

\usepackage{epsfig}
\usepackage{subcaption}
\usepackage{calc}
\usepackage{amssymb}
\usepackage{amstext}
\usepackage{amsmath}
\usepackage{amsthm}
\usepackage{multicol}
\usepackage{pslatex}
\usepackage{apalike}
\usepackage{algorithm2e}
\usepackage{url} 
\usepackage{xurl} 
\usepackage[hidelinks,implicit=false]{hyperref}
\usepackage{tcolorbox} 
\usepackage{mdframed}  
\usepackage[bottom]{footmisc}
\usepackage{SCITEPRESS}     
\newtcolorbox{takeawaybox}[1][]{
  colframe=gray!75,      
  colback=white!10,       
  coltitle=black,        
  boxrule=0.5mm,         
  fonttitle=\bfseries,   
  title=#1               
}

\begin{document}

\title{Autonomous Legacy Web Application Upgrades Using a Multi-Agent System}

\author{
    \authorname{
        Valtteri Ala-Salmi\sup{1}, Zeeshan Rasheed\sup{1}, Abdul Malik Sami\sup{1}, Zheying Zhang\sup{1}, Kai-Kristian Kemell\sup{1}, Jussi Rasku\sup{1}, Shahbaz Siddeeq\sup{1}, Mika Saari\sup{1}, and Pekka Abrahamsson\sup{1}
    }
    \affiliation{\sup{1}Faculty of Information Technology and Communication Science, Tampere University, Finland}
    \email{\{valtteri.ala-salmi, zeeshan.rasheed, malik.sami, zheying.zhang, kai-kristian.kemell, jussi.rasku, shahbaz.siddeeq, mika.saari, pekka.abrahamsson\}@tuni.fi}
}


\abstract{The use of Large Language Models (LLMs) for autonomously generating code has become a topic of interest in emerging technologies. 
As the technology improves, new possibilities for LLMs use in programming continue to expand such as code refactoring, security enhancements, and legacy application upgrades. Nowadays, a large number of web applications on the internet are outdated, raising challenges related to security and reliability. Many companies continue to use these applications because upgrading to the latest technologies is often a complex and costly task.
To this end, we proposed LLM based multi-agent system that autonomously upgrade the legacy web application into latest version. The proposed multi-agent system distributes tasks across multiple phases and updates all files to the latest version. To evaluate the proposed multi-agent system, we utilized Zero-Shot Learning (ZSL) and One-Shot Learning (OSL) prompts, providing the same instructions for both. The evaluation process was conducted by updating a number of view files in the application and counting the amount and type of errors in the resulting files. In more complex tasks, the amount of succeeded requirements was counted. The prompts were run with the proposed system and with the LLM as a standalone. The process was repeated multiple times to take the stochastic nature of LLM's into account. The result indicates that the proposed system is able to keep context of the updating process across various tasks and multiple agents. The system could return better solutions compared to the base model in some test cases. Based on the evaluation, the system contributes as a working foundation for future model implementations with existing code. The study also shows the capability of LLM to update small outdated files with high precision, even with basic prompts. The code is publicly available on GitHub: \url{https://github.com/alasalm1/Multi-agent-pipeline}.}

\onecolumn \maketitle \normalsize \setcounter{footnote}{0} \vfill

\section{\uppercase{Introduction}}
\label{sec:introduction}

Generative Artificial Intelligence (GAI) has advanced rapidly in the previous years. With the network model transformer proposed in \cite{vaswani2017attention}, the previous computational limitations related to neural networks has been passed. Compared to earlier architectures like Recurrent Neural Networks (RNNs), transformers are capable of being parallelized, allowing tokens to be processed simultaneously in a self-attention mechanism \cite{vaswani2017attention}. This allows efficient scalable computation of generative models in a graphical/tensor processing units, leading their possible size to grow greatly. As the size of the generative models increases, their performance improves, as demonstrated in language models by \cite{kaplan_scaling_2020}.
Large Language Models (LLMs) with advanced transformer architectures, such as Generative Pre-trained Transformers (GPTs) \cite{radford_improving_2018} and Bidirectional Encoder Representations from Transformers (BERTs) \cite{devlin2018pre}, have emerged as sizable and influential models in natural language processing.

With LLM's improving as larger and better trained models are introduced, the question arises as to whether the models could be used to update existing applications. The internet hosts a huge number of websites that contain deprecated components, which may affect their functionality and compatibility with modern standards. According to \cite{hohlfeld_our_2021}, 95\% of the analyzed 5.6 million web applications had at least one deprecated component. Updating an application takes notable resources which grows with the technical debt and eventually it becomes deprecated. The application owners are often stuck with the application as it is crucial to the business and updating is not possible without a big investment \cite{ali_addressing_2020}. 
Due to companies' hesitancy to undertake complex and potentially cost-ineffective operations, challenges arise regarding the safety \cite{smyth_penetration_2023,sami2024early}, usability \cite{ali_addressing_2020}, \cite{rasheed2024timeless}, and compatibility \cite{antal2016transforming}, of the web application.

In this study, we propose a multi-agent system to conduct a complex set of operations in phases for updating legacy application files. Each agent is assigned specific tasks, working collaboratively to accomplish the overall objective. For instance, a verification agent gives feedback for every executed implementation phase to ensure that phase's validity. Another agent called the finalization agent will then complete the phase if the verification agent sees a reason for it. The results of the proposed system are then tested using Zero-Shot Learning (ZSL) and One-Shot Learning (OSL) prompts \cite{brown_language_2020} to assess the system's suitability for the process and the overall capability of current LLMs to implement the update.

The comparison and evaluation of the system and prompts were conducted by using them to update five view files' web framework compatibility in a legacy web application. A ZSL prompt was used in comparison with four files and an OSL prompt with two files against the system. In the evaluation, the same error was counted once, giving a suitable metric to evaluate the performance of the operation. In more complex tasks, the completed requirements were counted to evaluate problem-solving in more challenging scenarios. The results show that the system, on average, updated a file with 0.46 different errors higher when compared against the best prompt implementation. With the requirements the results were varied with one task the system completed better and one task with a lower performance than the compared prompt. The evaluation results are publicly available for validating the study \cite{tampere_university_2024_14423728}.

Below our contribution can be summarized as follows:

\begin{itemize}
    \item The proposed system is designed to autonomously update legacy web applications to the latest version using a multi-agent system.

    \item The system was evaluated using an existing legacy web application and updating view files belonging to it. Then depending on the task, errors or fulfilled requirements are counted. The system had 0.46 more errors on average and varying performance in the requirements depending on the task.

    \item The system was compared to the standalone prompts giving perspective of the systems functionality and overall capability of LLM to implement code updating tasks.

    \item We publicly released the evaluation results dataset to access all the collected data for validating our study \cite{tampere_university_2024_14423728}.
\end{itemize}


The rest of the paper is organized as follows: Section \ref{Background Study} presents a background study on code generation using LLMs. Section \ref{Research Method} explains the methodology of this paper, followed by the results in Section \ref{Results}. Section \ref{Discussion} discusses the implications and future directions of the results, and the study concludes in Section \ref{Conclusion}.


\section{Background}
\label{Background Study}
In this section, we discuss the background of existing research. Section \ref{AA} examines prior agent systems developed for code generation, while Section \ref{BB} addresses the challenges associated with code generation using LLMs. Section \ref{CC} examines the challenges in upgrading legacy applications.

\subsection{Code Generation Using AI Agents}
\label{AA}
Lately, various studies have implemented different agent systems to improve code generation in LLM's \cite{rasheed2023autonomous}, \cite{rasheed2024ai}. This subsection investigates different implementations and their observed benefits compared to baseline code generation using LLM. 

The ability of using a self-feedback loop to improve code output has been observed in Madaan \textit{et al}. \cite{madaan2024self} with implementation of SELF-REFINE. It consists of a base prompt, a feedback prompt and a refiner prompt in which feedback system gives feedback of the base result and refiner improves it iteratively. SELF-REFINE had 8.7\% improvement in GPT-4 model in code optimization \cite{madaan2024self}.

In another implementation called Reflexion studied in Shinn \textit{et al}. \cite{shinn2024reflexion} an agent was connected to an evaluator which gave numeric evaluation of the agent output in the given operation. This was then processed in a self-reflection system which added textual reflective analysis in the agents memory to be used in the next outputs. Reflexion could perform in the HumanEval benchmark, a dataset containing Python programming, with the result of 91.0\%  \cite{shinn2024reflexion}.

One way of improvement has been testing the produced code to get a real feedback whether the produced code works. In Huang \textit{et al}. \cite{huang_agentcoder_2024}  system called AgentCoder was proposed. In the system one agent was tasked to produce code, the second one was tasked to invent test cases to the code and the last agent executed the test cases and provided results from the test to other agents. A feedback loop based on a real testing data achieved 32.7\%  better compared to ZSL prompt of GPT-4 \cite{huang_agentcoder_2024}. In HumanEval it obtained similar result as Reflexion, resulting in 91.5\% score \cite{huang_agentcoder_2024}.

Another implementation based on testing was created in Zhong \textit{et al}. \cite{zhong_debug_2024} named Large Language Debugger (LDB). In the system an agent generated program was divided into control blocks and then individual test cases were created and executed to each block by the agent-model. With LDB connected to Reflexion framework the resulting HumanEval percentage was measured in 95.1\% \cite{zhong_debug_2024}. 

Based on these studies, both self reflection of an agent and testing of the produced code did result in a remarkable rise of quality in code output in LLMs. The combination of both in the same system did give the highest result, making them possibly mutually reinforcing in code generation quality.

\subsection{Challenges in LLM Generated Code}
\label{BB}
The code generated by LLM has challenges that have been recognized in the literature. During background study, following challenges were found in three studies focusing on recognizing them: 

\begin{takeawaybox}[Challenges in code generation using LLM]
\scriptsize
\begin{itemize}
  \item The quality of the produced code decreases with the length of the code and the difficulty of the task \cite{liu2024refining,dou_whats_2024,chong_artificial-intelligence_2024}. 
  \item LLM does not always consider the requirements of the user, or those which are needed for reliable and safe code \cite{liu2024refining,dou_whats_2024,chong_artificial-intelligence_2024}.
  \item LLM can find problems in the code that do not exist when asked to give feedback from it \cite{liu2024refining,chong_artificial-intelligence_2024}.
\end{itemize}
\end{takeawaybox}

 The quality decrease of the LLM code by length and difficulty was found in two studies. In \cite{liu2024refining} LLM's ability to solve different Python and Java coding tasks was tested. Based on the study findings the probability of returned code working correctly decreases with harder code tasks and the length of the produced code. In another study \cite{dou_whats_2024}, different programming tasks were generated by different LLMs and evaluated by syntax, runtime, and LLM's functionality errors. The study found an increased failure rate with more code lines, code complexity, and required API calls to produce code. In \cite{chong_artificial-intelligence_2024} the security of LLM-generated code was evaluated. The findings show that creating a memory buffer correctly in complex tasks, for example, the multiplication of two floats had a much lower success rate of 1.5\% than in an easier task like subtracting a float from an integer with a success rate of 50.1\% \cite{chong_artificial-intelligence_2024}.

LLM also has problems producing reliable, safe code and occasionally failing the user request \cite{rasheed2024large}, \cite{10.1007/978-3-031-78386-9_20}. In Liu \textit{et al}. errors found in the code were mostly related to wrong outputs (27\%) and badly styled and lowly maintainable code (47\%) while runtime errors were much lower (4\%) \cite{liu2024refining}. The badly styled and lowly maintainable code included categories like a redundant modifier, ambiguously named variables, and too many local variables in a function or method. Dou \textit{et al}. found that the LLM's functionality is the most notable reason for bugs in the code, particularly misunderstanding of the provided problem and logical errors in the code \cite{dou_whats_2024}. This included, for example, failed corner case checking, undefined conditional branches, and a complete misunderstanding of the provided problem. Chong \textit{et al}. compared LLM code to human-generated code in 220 files. The study found that while LLM generates fewer lines of code, the code lacks defensive programming that exists in the human-written code \cite{chong_artificial-intelligence_2024}. Additionally, an SHA generation algorithm was provided by LLM as a faulty version while AES and MD5 succeeded, making the completion of similar tasks unreliable \cite{chong_artificial-intelligence_2024}.

The studies found that while a feedback loop can improve code generation it can also cause additional errors to the code. Liu \textit{et al} found that giving feedback improves produced code up to 60\% but with a possibility of additional errors added to the code \cite{liu2024refining}. Chong \textit{et al}. found that the LLM can generate new security problems in the code by a feedback loop and not just remove them, especially if the file does not contain problems in the first place.  

When compared to the studies of multi-agent systems, it seems that observed challenges can be improved with the help of a multi-agent system. The observed improvement of self-feedback \cite{liu2024refining,chong_artificial-intelligence_2024}, has been implemented in \cite{madaan2024self,shinn2024reflexion,huang_agentcoder_2024,zhong_debug_2024}, resulting in improvement in the code evaluation metrics despite the risks of additional errors. Additional improvement using test results in \cite{huang_agentcoder_2024,zhong_debug_2024}, can also be seen as a way to solve LLM's problems in functionality as observed in \cite{dou_whats_2024}. As multi-agent systems have been shown to add value to code generation in LLM by being able to address some of its challenges, the proposed multi-agent pipeline is expected to add value in updating code in software.  

\subsection{Challenges in Upgrading Legacy Applications}
\label{CC}
Legacy systems are outdated implementations by used technologies and programming languages with hardware, software, and other parts of the system being possibly obsolete \cite{sommerville_software_2016}. When it comes to strategies for dealing with legacy systems the possibilities are disposing of the system, keeping the system as such, and re-engineering or replacing components in the legacy system \cite{sommerville_software_2016}. In this study, the focus is on re-engineering and replacing components on the application side with the help of a multi-agent system. Below is a summary of found challenges in legacy applications based on case studies:

\begin{takeawaybox}[Challenges in updating legacy applications]
\scriptsize
\begin{itemize}
  \item The programmers can have knowledge gaps in either old or new technologies of a legacy application \cite{de_marco_cobol_2018,fritzsch_microservices_2019}. 
  \item Identifying updated components' input, output, and code functionality is a demanding, time taking task in legacy applications \cite{de_marco_cobol_2018,vesic_framework_2023}.
  \item 
  Breaking the updating process down into smaller parts and successfully managing them is a challenge in legacy applications \cite{de_marco_cobol_2018,fritzsch_microservices_2019}. 
\end{itemize}
\end{takeawaybox}

First, programmers might have knowledge gaps in technology either in the original legacy application or the new version. As mentioned by \cite{de_marco_cobol_2018}, a legacy mainframe application was migrated to Linux servers with changes to the database and a transition in programming language from COBOL to Java. With feature development, the paper mentions that the COBOL team has challenges with Java programming language and vice versa, causing a knowledge gap between the teams \cite{de_marco_cobol_2018}. In Fritzsch \textit{et al}. 14 legacy applications in different stages of migration to microservices were analysed \cite{fritzsch_microservices_2019}. When it came to the recognized challenges in the migration, the lack of expertise was the shared first cause as knowledge of microservice architecture was not high enough with the developers \cite{fritzsch_microservices_2019}.

Second, a component of a legacy application has a challenging task to correctly recognize correct input, output, and code functionality. In De Marco \textit{et al}. the testing phase of the migration caused one-year delay for the project. The reason for this included a lack of high-level tests that made recognizing input, output, and inner functionalities of a component a time-taking task \cite{de_marco_cobol_2018}. Additionally, obsolete code took time to correctly identify its functionality inside a component \cite{de_marco_cobol_2018}. In  \cite{vesic_framework_2023} a framework for legacy system evaluation was presented with an analyse of an existing legacy system. During the analysis the studied information system of water and sewerage disposal company showed multiple problems. The system lacked proper documentation, lack of personnel and people with knowledge of the whole system, and poor software architecture \cite{vesic_framework_2023}. These problems shows that if the software side were updated, identifying of the component's behavior would be a complex task as in the \cite{de_marco_cobol_2018}.

Lastly, breaking the legacy system update into smaller tasks and managing them is a recognised challenge. In Fritzsch \textit{et al}. decomposition of the application was the second shared first reason for technical challenges in studied projects \cite{fritzsch_microservices_2019}. In De Marco \textit{et al}. the decomposited work packets were tried to be used to successfully forecast the project duration but failed due to differences in batch-orientation \cite{de_marco_cobol_2018}. 

As stated in section \ref{sec:introduction}, at least one component is considered obsolete in 95\% of web applications \cite{hohlfeld_our_2021}, making the problem relevant in the industry. The recognized challenges found in updating legacy systems could be solved with the help of an LLM. As LLM can be trained to have knowledge of various technologies \cite{brown_language_2020}, it could help with the knowledge gap of project developers. Additionally, LLM could analyse application components and save time by understanding the component's functionality. LLM could be used to divide and manage tasks of the project at different levels. Combined with the findings of the previous subsections, a multi-agent system that could provide solution to these challenges is a relevant option to conduct upgrades in legacy applications.


\begin{figure*}[!t]
\centering
\includegraphics[scale=0.2]{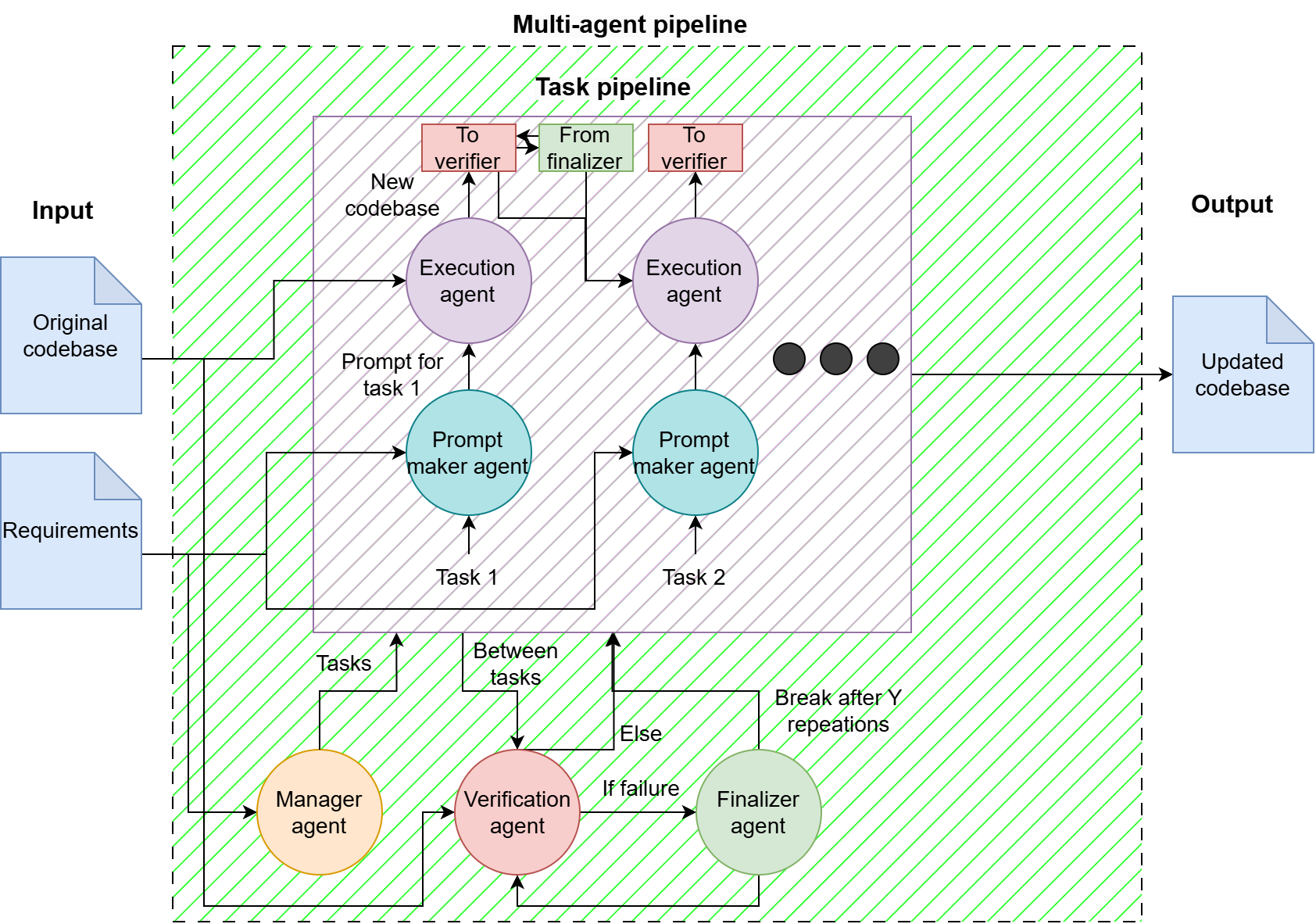}
\caption{Proposed system: Multi-agent pipeline for updating existing code}
\label{fig:fig_model}
\end{figure*}

\section{Research Method}
\label{Research Method}

In this section, the Research Questions (RQs) are first introduced, followed by a detailed explanation of the proposed system. Next, the evaluation subjects and objectives are presented. Finally, the evaluation process for the proposed system is described.

\subsection{Research Questions}\label{RQ}
In this study following two RQs are proposed:


\begin{tcolorbox}[colback=gray!2!white,colframe=black!75!black]
\textit{\textbf{RQ1.} How to utilize multi-agent system to update legacy project into latest by refactoring deprecated code?}
\end{tcolorbox}

The aim of RQ1 is to test the capability of a multi-agent system to autonomously upgrade legacy web applications. This objective focuses on assessing how effectively the system can modify and enhance outdated functionalities by refactoring deprecated code without manual intervention. 

\begin{tcolorbox}[colback=gray!2!white,colframe=black!75!black]
\textit{\textbf{RQ2.} How to validate the proposed multi-agent system?}
\end{tcolorbox}

The aim of RQ2 is to define a suitable metric to validate a multi-agent system meant for autonomously update deprecated code. This objective focuses on creating a metric and then using it to validate the proposed system.




\subsection{Proposed Multi-Agent System}
\label{proposed:sysmte}

The proposed multi-agent system is referenced from the CodePori multi-agent system described in \cite{rasheed2024codepori}. In the CodePori system, software operating in Python code are generated based on a given project description instructing the system, for example, creating a simple game or a face recognizer \cite{rasheed2024codepori}. The outcome is generated by a six agent-framework tasked to operate different software development team roles including a manager, developers, finalizer, and a verifier.

In the proposed system illustrated in figure \ref{fig:fig_model} the multi-agent system is designed to modify already existing code based on the user requirements. The system receives an original codebase and the requirements from the user as input. In the requirements the user specifies, what type of operations they want to commit to the code. For example, updating the code compatibility from version X to version Y or changing the libraries that the code uses. The system is composed of four units which are explained below:

\textbf{Manager Agent:} The manager agent receives the updating requirements from the user and is tasked to write them into manageable operations in a chronological execution order. The tasks are written in abstract level, which are later defined by the task pipeline where the list of tasks is send. The manager agent is asked once before sending the tasks to ensure that the tasks are in chronological order and overall related to the requirements.

\textbf{Task pipeline:} The task pipeline consists of prompt makers and execution agents illustrated in figure \ref{fig:fig_model} inside multi-agent pipeline. For every task the prompt maker agent creates an OSL prompt which is executed by execution agent. This creates a pipeline of OSL prompts that are executed sequentially to the given code. After every task the code is send for the verification agent for reviewing the completion of the task. When every task is completed in the pipeline it will give the updated code as an output for the user.

\textbf{Verification agent:} The role of the verification agent is to ensure that a task has been completed. It will analyse the new version of the code and estimate whether it satisfies the requirements. If the verification agent accepts the task, it will return the code for the new task in the pipeline. If there is something left in the task, it will send the code to the finalizer agent.

\textbf{Finalizer agent:} The finalizer agent makes changes to the updated code if the verification agent notices that the tasked operation has not been completed successfully. After executing changes, the finalizer agent returns the modified code back to verification agent for the next analyze. If the feedback loop between the verification agent and the finalizer agent exceeds a certain amount of interactions, the finalizer agent will send the code back to the task pipeline. This might happen, for example, if the verification agent starts to hallucinate and find problems that does not exist.

Expected improvements of the system compared to ZSL/OSL prompts when updating existing code are based on the following features:

\begin{itemize}

\item[1.] \textit{Self-division:} The division of code update into smaller tasks avoids particularly long prompts. As observed in \cite{liu2024lost}  information in long prompts is hard to access by LLM's especially in the middle of the prompt. As the whole task prompt contains instructions, the instructions in the middle will not be necessary completed, causing task to partially fail.

\item[2.] \textit{Self-feedback:} The output of every task will be analysed on the verifier agent and then possibly improved by the finalizer agent iteratively, generating a self-feedback loop. As found in the background studies,  self-feedback improves results in code generation \cite{liu2024refining,chong_artificial-intelligence_2024}. As updating the code to a new syntax and improving it's functionality keeps code logically same, the improvement is expected in the output.

\item[3.] \textit{Self-instructive:} Manually writing complex prompts to instruct LLM for updating existing code is a hard task for people that are not familiar with prompt engineering. In \cite{zamfirescu-pereira_why_2023} non expert prompt writers were studied, which found problems with over generalization and human interactive style of writing prompts resulting in bad performance. With self-instruction the user needs only to write the needed task without adding detailed instructions or examples. 

\end{itemize}

\subsection{Evaluation Subjects and Goal}

The evaluated legacy web application is built on CakePHP 1.2, which is a PHP based web framework with development started in 2005 \cite{cakephp_introduction_2022}. During this evaluation the main goal is to update files to version 4.5  of the web framework \cite{cakephp_cakephp_2023} with additional challenges that are related to each updated file. CakePHP 1.2 was released on 2008 \cite{koschuetzki_extra_2008} and CakePHP 4.5 on 2023 \cite{cakephp_cakephp_2023}, making the version gap between the versions 15 years. 

CakePHP is built based on Model-View-Controller (MVC) architecture \cite{cakephp_introduction_2022} which is a software design pattern with controller acting as a mediator between a model and a view. Between versions 1.2 and 4.5 CakePHP has got a multiple changes in the design and syntax. For example, in CakePHP 3.0 the Object–Relational Mapping (ORM) was re-built \cite{cakephp_30_2024} and the required version of PHP was raised to 7.4 in the version 4.x of CakePHP \cite{cakephp_installation_2024} from the original PHP version of 4 and 5 \cite{cakephp_introduction_2022}.

The comparison between the proposed system and the alternatives was conducted on a module consisting 5 view files, which belongs to the functionalities of an already updated controller file in the legacy web application. The web application in this study is an electronic dictionary described in \cite{norri_digitization_2020}. The dictionary is a search and editing tool for a Postgres database, which contains informative data on medieval medical English vocabulary. The component of five files form a part of the  dictionary where medieval medical variants can be searched based on different features like name, original language and wildcards \cite{norri_digitization_2020}. The files, their size and any recognized challenges are seen in the table \ref{tab:evaluated_files}.

   

\begin{table}[htbp]
\centering
\caption{Evaluated files}
\resizebox{\columnwidth}{!}{%
\begin{tabular}{|c|c|p{5cm}|}
\hline
\textbf{File} & \textbf{LOC} & \textbf{Challenges} \\ \hline
View A & 25 & Changed data format from Controller \\ \hline
View B & 57 & Changed data format from Controller \\ \hline
View C & 190 & Contains JavaScript and PHP \\ \hline
View D & 19 & Ajax form to be updated to JQuery \\ \hline
View E & 118 & 4 helper functions must be replaced \\ \hline
\end{tabular}%
}
\label{tab:evaluated_files}
\end{table}

View A is tasked to show quotes that are connected to a specific variant. View B shows searched variants that are part of a certain quote and uses the functionalities of View D and E. In both View A and B, a requirement is to take changed data format into account as the Controller in the version 4.5 uses ORM resultsets introduced in CakePHP 3 \cite{cakephp_retrieving_2024}. However, unlike in the normal case where reference is done by object syntax, the reference must be done by array syntax with first field-name starting in a capital letter.

View C is a dynamic list that shows variants based on their first letter and the language of origin. View C contains both PHP and JavaScript code to add challenge to updating process. Unlike in View A and B, the ORM objects are now accessed with the normal notation.

View D is an Ajax form used in View B which is needed to be remade with jQuery. Therefore in addition for updating the CakePHP syntax the architecture needs to be updated as well. View E file is an element view file which is made to highlight a search world of the search results in the view B file. Besides of the highlight functionalities, the view has been added four helper functions that are wanted to be replaced with a modern library implementations.

\subsection{Evaluation Process}
\label{EP}

The proposed system was tested against a ZSL and, when needed, against an OSL prompt to update the view files. In ZSL, the prompt is defined to not include any examples of the task, whereas a OSL prompt includes exactly one example of the given task \cite{brown_language_2020}. An OSL prompt is typically used when a task requires a custom example to guide the model's behavior, especially in cases where a ZSL prompt does not produce the desired results. The use of prompts as a metric to evaluate LLM techniques was explored by Ouedraogo \textit{et al}. \cite{ouedraogo_large-scale_2024}, where ZSL and OSL prompts were compared with different reasoning methods for LLMs.

In a ZSL prompt the GPT-model is asked to update a view file without any examples. An OSL prompt includes instructions for updating the file with an example given. The loopback in the system is set with the maximum of two iterations. The system and the compared alternatives were ran with the ChatGPT 4o-mini model \cite{openai_gpt-4o_2024}. 

The test was repeated multiple times for every prompt/file to take the stochastic nature of the LLM's into account \cite{brown_language_2020}. When a view file was tested, the connected controller file was in the new version of the web framework.

The View files A, B and C were evaluated by different errors in the code. The evaluation was done manually with both static and dynamic testing used to find errors from the updated file. The errors found were divided into following categories:
\begin{itemize}

\item[1.] \textit{Fatal Errors:} Errors that causes the file to not run, for example, syntax errors.
\item[2.] \textit{Runtime Errors:} Errors that do not prevent running the file but are encountered during the usage of the file.
\item[3.] \textit{Content Errors:} The feature works but its functionality is different than in the original file.
\item[4.] \textit{Missing/Additional features:} The updated file is missing or have additional features not existing in the file
\item[5.] \textit{Failed generation:} If the updated file has more than 7 different type of errors or the answer does not contain the code the generation is considered as failed and the evaluation is stopped.

\end{itemize}

The same error caused by the same mistake was counted only once to ensure that recurring syntax errors do not disproportionately affect the comparison with other types of errors. From the results a Standard Deviation (SD) was calculated to measure similarity of the code generation errors across the repetitions. Additionally the average duration of the request was counted along with the Lines of Code (LOC) in the updated files. The reasoning behind counting LOC is to analyse whether there is difference with the length of produced code between a OSL/ZSL prompt and the system. 

For the View D and E where the entire code is refactored and based on complex requirements, the evaluation process was decided to be ranked based on the requirements it passes. For every correctly working requirement it a value of 1 was given, and if not, value of 0 was given. The evaluation is conducted manually as the error counting.

The features for View D are following:

\begin{takeawaybox}[Requirements for the view D form update]
\scriptsize
\textbf{Requirements of the updated form:}
\begin{itemize}
  \item The form sends data and receives results as expected.
  \item Dropdown menus show correct suggestions.
  \item The send button and dropdown objects work as intended.
\end{itemize}
\end{takeawaybox}

Referring to jQuery file as a URL or a local file are both accepted. Also data name sent to controller is slightly different this is also disregarded from error evaluation. Besides these, no additional errors are fixed in the evaluation.

The results of View E are evaluated similar to the View D along with the number of replaced functions with library implementations. The requirements for View E are following:

\begin{takeawaybox}[Requirements for the view E highlight update]
\scriptsize
\textbf{Requirements of the highlighting feature:}
\begin{itemize}
  \item The highlighting works in normal case with no special characters or wildcards.
  \item Highlighting words with special medieval English letters.
  \item Highlighting works with wildcards with the highlight only to the part of word where the wildcard was placed in the search query.
\end{itemize}
\end{takeawaybox}

If some functions were not replaced, they are used in the file with added guard to not re-declare them in the testing process, but otherwise left as they were given by the test subject.

\section{Results}
\label{Results}
In this section, we present the results of our proposed system. The results related to the proposed system are provided in Section \ref{RQ1}, while the validation results are provided in Section \ref{Validation RQ2}.

\subsection{Suitability of Proposed System for Updating a Deprecated File (RQ1)}\label{RQ1}

Updating View A was first attempted with a ZSL prompt. The prompt to address this problem was written in the following format: 

"Update CakePHP view file from version 1.2 to version 4.5. 

Requirement: \$quotes is an ORM\textbackslash ResultSet made of arrays accessed with ['Fieldname']['fieldname'] and must be accessed so in the updated file."

This was added to a framework that included the updated code and a request to only return the updated code. The updating process were run ten times. After the run the resulted errors were counted in the files. The prompt returned the updated file in the average of 1.6 different errors and in average of 2.3 seconds. The prompt failed in all ten times to use ['Fieldname']['fieldname'] instead using format ['fieldname']['fieldname']. In a total of six times the prompt failed to access correctly to the ORM object resulting in a fatal error.

Based on these remarks an short OSL prompt was created trying to address the resulted problems in the following format:

"Update CakePHP view file from version 1.2 to version 4.5. Requirement: \$quotes are ORM\textbackslash ResultSets made of arrays accessed with ['Fieldname']['fieldname'] and must be accessed so in the updated file. Example: old syntax: ['apple']['lemon'] new syntax: ['Apple']['lemon']. Use function first() instead of [0] to the ORM object."

Used in the same framework, the results showed improvement with 0.4 of average amount of errors in the file. The file was updated correctly in total of seven times out of ten. As this is relatively low average, the prompt was kept such with the exception of the last sentence removed and tested with a more complex View B file. The results with the OSL prompt were once again impressive with average amount of errors being 0.7 with the file correctly updated four times.

The ZSL and OSL prompt were next compared with the proposed system. Based on the testing of the prompts, the requirement file was written into following format:

"Requirement1: Update whole CakePHP view file from version 1.2 to version 4.5. 

Requirement2: ORM Arrays must be accessed with array style syntax ['Fieldname']['fieldname'] with the first fieldname starting with a capitalized letter and the second only with lowercase letters. Use first() when referring to first member in the array."

The testing was conducted 10 times with the view files A and B with the last sentence removed. The system returned view A with an average of 53.2 seconds and 0.6 different errors. One of the runs was a statistical outlier with the return time of 165 seconds. The system succeeded four times to return the file correctly. With View B the system returned it with average return time of 60 seconds and 1.0 different errors on average. Total of 3 times the system returned the file correctly.

View C was updated using the system and a ZSL prompt. As the ORM objects are now accessed with the normal notation, therefore, prompt could be simplified notably and ZSL was determined to be enough for the comparision. Following ZSL prompt were used in the updating process:

"Update whole cakePHP view file from version 1.2 to version 4.5. Use ORM access with \$variant with direct access to name and id."

With repetition of 10 times with the prompt in the testing framework, the file was returned with an average of 10.7 seconds and average of 0.3 different errors. In total of seven times the prompt returned a fully correct file.

With the system, the updating process was conducted with the following requirements:

"Requirement1: Update whole cakePHP view file from version 1.2 to version 4.5. 

Requirement2: Use ORM access with \$variant with direct access to name and id."

The system performed poorer than a ZSL prompt with an average of errors 1.22 in 67.8 seconds. One generation did fail with the full code not being generated and is not included in the error averages. 

The values of the evaluation has been collected in the Table \ref{tab:small_files}. Based on the results, the proposed system and OSL/ZSL prompt are both capable of updating an deprecated code file with a high precision. However, OSL/ZSL prompt seems to perform slightly better in the evaluated files, especially in the View C. Average lines of code (ALOC) between methods were similar in size.

\begin{table}[h]
 \renewcommand{\arraystretch}{1.3}
 \caption{Evaluation of updated view files A, B and C}
 \label{tab:small_files}
 \centering
 \resizebox{\columnwidth}{!}{%
 \begin{tabular}{|c||c|c|c|c|c|}
 \hline
 \bfseries File & \bfseries Method  &\bfseries Different errors&\bfseries SD&\bfseries ALOC&\bfseries Time (s)\\
 \hline\hline
  View A & ZSL &1.6&0.490&22&2.3\\
  \hline
  View A & OSL &0.4&0.663&22&5\\
 \hline
  View B &OSL&0.70&0.640&57&7.4\\
  \hline
    View C & ZSL &0.3&0.459&160&10.7\\
   \hline
   View A &Syst.&0.60&0.490&24&53.2\\
   \hline
   View B &Syst.&1.0&0.850&60&60.4\\
   \hline
    View C &Syst.&1.22&0.786&164&70.9\\
   \hline
   
   \hline
 \end{tabular}%
}
 \end{table}


\begin{figure*}[!t]
\centering
\includegraphics[scale=1]{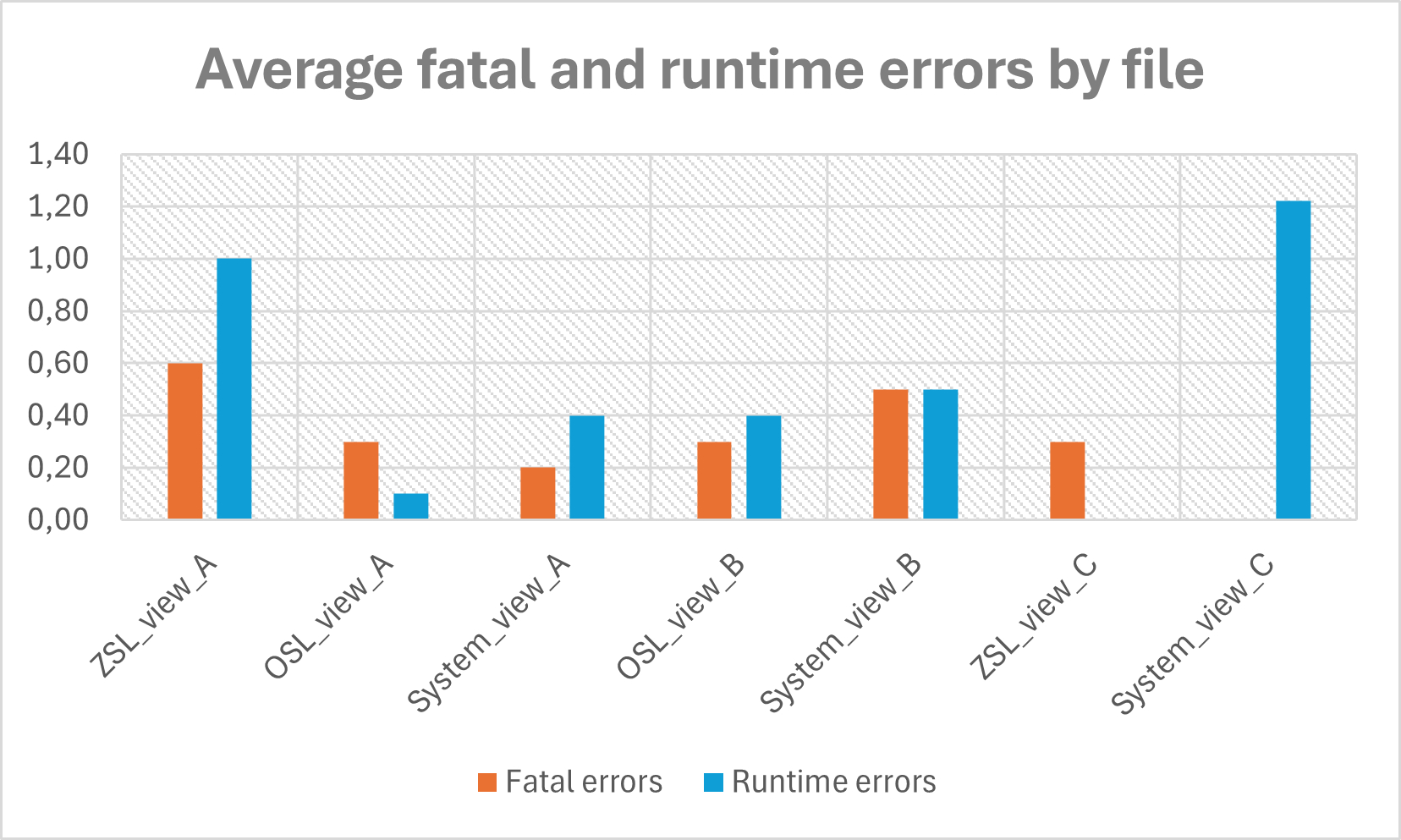}
\caption{Found error types by method and file}
\label{fig:found_errors}
\end{figure*}

During the evaluation the different error types were counted and are shown in figure \ref{fig:found_errors}. Only fatal and runtime errors were found from the generated files. Interestingly the proposed system did generate lesser runtime errors in View A and C, however it did generate more runtime errors in every file. Particularly in View C all ZSL errors were runtime and all proposed system errors were runtime.

\subsection{Validation of the Proposed System (RQ2)}
\label{Validation RQ2}





The validation process for the  harder tasks for View D and E was conducted with a ZSL prompt and the system. Following ZSL prompt was made to task the update in View D: 

"Update CakePHP version 1.2 ajax form into CakePHP 4.5 version with jQuery architecture including jquery-3.6.0.min file. Make the jQuery implementation fully functional version with dropdown updated every time the letter is written to it"

With the system the prompt was split into two requirements by sentences. The test was repeated with both method 10 times and the files were evaluated based on the requirements and added to Table \ref{tab:table_d}. The ZSL was able to fulfill the request with a value of 0.5 with once correctly providing completely functional form. The system had value of 0.9 with the view, twice giving the fully correct form. The system created on average twice as large code file compared to the ZSL implementation. 

\begin{table}[ht]
 \renewcommand{\arraystretch}{1.3}
 \caption{Evaluation of view D based on passed requirements by average}
 \label{tab:table_d}
 \centering
 \resizebox{\columnwidth}{!}{%
 \begin{tabular}{|c||c|c|c|c|c|}
 \hline
 \bfseries Method & \bfseries Regt 1  &\bfseries Regt 2&\bfseries Regt 3\bfseries&\bfseries Total&\bfseries ALOC\\
 \hline\hline
  ZSL & 0.2 &0.1&0.2&0.5&42\\
  \hline
   Syst. (2 tasks)& 0.5 &0.3&0.2&0.9&86\\
   \hline
 \end{tabular}%
}
 \end{table}





The ZSL prompt for View E was formed as following with the model divided into two requirements by sentences:

"Update CakePHP element view file from 1.2 to 4.5. Replace functions in the element view file by ready made PHP libraries or update them if not found"

The results are collected in the table \ref{tab:table_e}. The ZSL prompt had average requirement value of 1.8 and average of 3.2 replaced functions (RF) with five files passing all requirements. With the system dividing the prompt in the two tasks resulted average requirement value of 0.5 with 2.7 RF with no completely passing version. This was repeated with the system running with only one task resulting total requirement value of 1.6 and value of 1.7 in RF with four files passing the requirements fully.

\begin{table}[h]
 \renewcommand{\arraystretch}{1.3}
 \caption{Evaluation of view E based on passed requirements and replaced functions by average}
 \label{tab:table_e}
 \centering
 \resizebox{\columnwidth}{!}{%
 \begin{tabular}{|c||c|c|c|c|c|c|}
 \hline
 \bfseries Method & \bfseries Regt 1  &\bfseries Regt 2&\bfseries Regt 3\bfseries&\bfseries Total&\bfseries RF&\bfseries ALOC\\
 \hline\hline
  ZSL & 0.7 &0.6&0.5&1.8&3.2&42\\
  \hline
   Syst. (1 task)& 0.6 &0.6&0.4&1.6&1.7&57\\
   \hline
   Syst. (2 tasks)& 0.3 &0&0.2&0.5&2.7&52\\
   \hline
 \end{tabular}%
  }
 \end{table}

The results show that the system performed much weaker compared to the ZSL prompt when divided into two subtasks. When the system was run on a single task, the results closely aligned with the requirements; however, the ZSL prompt successfully replaced more functions with library alternatives. Most issues encountered during file updates were related to references to non-existent CakePHP functions, highlighting hallucinations commonly found in code generated by LLMs \cite{dou_whats_2024}.

Based on these tests, the proposed system does not perform notably better compared to ZSL in more complex problem solving tasks but in some settings might perform much worse compared to a ZSL prompt. Overall, the results indicates ability of LLM to solve complex coding tasks like library replacements and transforming code to work in different library architectures. The evaluation results are publicly available for further validation \cite{tampere_university_2024_14423728}.

\section{Discussion}
\label{Discussion}

During this study files belonging to an existing deprecated web application were updated using GAI. The results shows that multi agent systems are capable of updating small deprecated web application files with a high precision with low rates of error and a low mean deviation of errors between the generations. The multi agent system could provide completely working versions of code in every studied task that required, for example, library replacements.

The observed ability of LLM to generate not only new code but to update existing one has an potential impact for the software industry. As notable part of the industry is related for upkeeping existing applications, the possibility of automatising the upkeep process by at least partially by artificial intelligence, will result in much faster and cheaper process of the application updating. Estimations of code maintenance has been ranged, for example, at 85-90\% \cite{erlikh_leveraging_2000} and 40-80\% \cite{davis_97_2009}. With the automation the percentage would be expected to be lowered considerably, freeing resources of software companies.

The question also rises for LLM's capability of transforming existing code across coding languages. If the context can be kept enough, a system designed to such task could translate code across coding languages similar to human languages. This already have studies \cite{pan_lost_2024,eniser_towards_2024},  that shows different methods to translate code in various languages. However the translation percentage did not in either of those studies rise above 50\%, indicating that more advancements are still needed in the field to reach necessary level of quality.

A multi-agent system called multi-agent pipeline was proposed in this study for completing a code update in sequences with a self-feedback loop. The results shows that ZSL/OSL prompt produces usually better results compared to the proposed system. There are possible multiple reasons for this:

\textbf{Telephone game:} Telephone game is a play where information is passed in a chain for one player to another. The longer the game continues more distorted the original message becomes. With a chain of agents the possibility of code logic starting to cumulatively change is a possible reason for lower results. Despite the verifier agent checking the results compared to the original code, the possibility of distorted code is still existing if it goes undetected by the verifier agent. Another risk is the hallucinations which added to the codebase can create faulty code which was encountered when updating the view E file.

\textbf{LLM reasoning skills:} The system used in testing 4o mini might not have the reasoning skills required to fulfill the tasks of more complex agents like the verifier or the manager agent. Testing of complex multi-agent roles needs to be repeated in a more advanced system and compare results to the existing ones to. For example, new LLM specialised for problem solving like GPT o1 \cite{openai_introducing_2024}  could make verification agent better performing with the potential ability in problem solving. 

\textbf{Prompt following:} As found in \cite{dou_whats_2024} the LLM has challenges to understand the given task from the prompt. With the proposed system the prompt following capabilities did not have notable improvement compared to the tested short prompts as seen with the fulfilled requirements. To address this the given prompt might need to be further processed with agents before sent to the task-pipeline to address problem of misunderstanding the given task.

Based on these hypothesis for the systems under-performance, alternative versions of the system are needed to be explored to investigate their effect on the performance. Overall the system ability to update files correctly and sometimes better than the OSL/ZSL makes the system as a foundation for better refined versions in the future. 

Besides of the system improvement other possible future work are recognized. Based on the background studies \cite{huang_agentcoder_2024,zhong_debug_2024}, having a test bench for code evaluation can have a positive impact on the outputted code. As future work evaluating a multi-agent system with a test bench might provide way to improve the results. 

\subsection{Threats to Validity}

This paper recognizes limitations and risks associated with the methods used in this study. First, the evaluation being limited to five files due to manual evaluation, risks incomplete comparison between methods and possible mistakes with the classification of different errors. Also validating the results in other studies is more challenging without a recognized metric like HumanEval in studies \cite{shinn2024reflexion,huang_agentcoder_2024,zhong_debug_2024}.

When it comes with the used prompts in this study, it is important to take into account that different prompts could give potentially different results. As prompt engineering with multi-agent models is not a subject in this study a differently crafted prompt could give a better result. This is also a possibility with the OSL/ZSL prompts.

There are also risks with the validity of the proposed system. A possible error in the system could give false results, for example, if one of the agents has poor instructions to operate. Lastly, It is important to note that the results of the studied files can not be generalized in other use scenarios and more extensive survey is needed to evaluate code updating abilities across different frameworks, coding languages and use cases. 

\section{Conclusion}
\label{Conclusion}

In this study the capabilities of LLM for updating existing code were investigated using an GPT model in a deprecated web application. The results shows that LLM's are capable of updating small files with high precision using short ZSL and OSL prompts. The study proposed an multi-agent system called multi-agent pipeline to improve code to update results of the LLM code output.

The evaluation of the system showed that while the system is capable of mostly update files like the alternative ZSL/OSL prompts, it did not offer increase for performance and, in some cases, under performed compared to the alternatives. The proposed system however offers a foundation for future multi-agent systems designed for code updating. For the future improvements multi-agent system needs to address challenges related in updating of existing code. 


\bibliographystyle{apalike}
{\small
\bibliography{example}}



\end{document}